\documentclass[twocolumn,twoside,11pt]{article}
\begin{document}

 \newcommand\ceil[1]{{\lceil#1\rceil}}
 \newcommand\floor[1]{{\lfloor#1\rfloor}}
 \newcommand\cf{{\em cf\/}} \newcommand\eg{{\em e.g.,\ }}
 \newcommand\etc{{\em etc}} \newcommand\ie{{\em i.e.,\ }}
 \newsavebox\copyrt\sbox\copyrt{\rule[-8pc]{0pc}{0pc}}
 \newcommand\qed{\rule{1em}{1.5ex}}

 \newtheorem{thr} {Theorem} \newtheorem{lem} {Lemma}
 \newtheorem{dfn} {Definition} \newtheorem{ntn} {Notation}
 \newtheorem{prp} {Proposition} \newtheorem{cor} {Corollary}
 \newtheorem{rem}{Remark}\newtheorem{nt}{Note}\newtheorem{cmt}{Comment}

\newcommand{\cast}[1]{{$#1$-cast}}
\newcommand{\clique}[1]{{$#1$-clique}} \newcommand{\set}[1]{{$#1$-set}}
\newcommand\byz{\mbox{\rm\bf Byz}}

 \title {\vspace*{-3pc} Byzantine Agreement with Faulty Majority using
Bounded Broadcast}
 \date{} \author {Jeffrey Considine, Leonid A.~Levin\thanks{Supported by
NSF grants CCR 9820934, 0311411}, David Metcalf
 \\ Boston University\thanks {Leonid A. Levin (Lnd at bu.edu),\hfill
Computer Sci. dept., 111 Cummington St., Boston, MA 02215.
 %\hfill\mbox {Last revised: \today}
 }}\maketitle

 \begin{abstract} Byzantine Agreement introduced in [Pease, Shostak,
Lamport, 80] is a widely used building block of reliable distributed
protocols. It simulates broadcast despite the presence of faulty parties
within the network, traditionally using only private unicast links.
Under such conditions, Byzantine Agreement requires more than 2/3 of the
parties to be compliant. [Fitzi, Maurer, 00], constructed a Byzantine
Agreement protocol for any compliant majority based on an additional
primitive allowing transmission to any two parties simultaneously. They
proposed a problem of generalizing these results to wider channels and
fewer compliant parties. We prove that $2f<kh$ condition is necessary
and sufficient for implementing broadcast with $h$ compliant and $f$
faulty parties using \cast{k} channels. \end{abstract}

\section{Introduction}

Broadcast primitives play a special role in multiplayer game theory as
an integral component in the fault-tolerant implementation of game
protocols. Given a compliant majority, broadcast and private channels
are sufficient to simulate any multi-party computation [Rabin, Ben-Or,
89], based on [Goldreich, Micali, Wigderson, 87], and [Ben-Or,
Goldwasser, Wigderson, 88]. With additional primitives, such as private
and oblivious transfer channels, even a majority of faulty parties can
be tolerated [Beaver, Goldwasser, 89], [Goldwasser, Levin. 90].

Since reliable hardware solution is a strong assumption, Byzantine
Agreement protocols simulate broadcast on networks with faulty parties.
Given only private channels, Byzantine Agreement is possible if and only
if faulty parties are in a $<1/3$ minority ([Pease, Shostak, Lamport,
80]). For this reason, protocols tolerant to more faults generally
assume broadcast as a primitive.

Various hardware assumptions and communication goals were studied in the
literature. For instance, [Angluin 80], [Goldreich, Goldwasser, Linial
91], [Franklin, Yung 95] and other papers studied problems of private
communication on an incomplete broadcast network. [Franklin, Wright 98]
showed that such a network with $p$ disjoint paths from sender to
receiver could tolerate $<p/2$ faulty parties. [Wang, Desmedt 01] showed
that $<p$ faulty parties could be handled with probabilistic
reliability.

The broadcast primitive is rather special in that, unlike other common
primitives, it involves an unlimited number of parties. This suggests
exploring the power of a limited version of broadcast, with a constant
number of recipients. Assuming any compliant majority, [Fitzi, Maurer,
00] used a 3-party broadcast primitive to simulate full broadcast. They
asked what fraction of compliant parties would be required given wider
broadcast primitives. This is especially interesting in view of results
({e.g.} \cite{BG,GL}) that convert arbitrary protocols into equivalent
ones with added tolerance to any faulty majorities, assuming the
availability of broadcasts and two-party primitives, such as oblivious
transfer and private channels. We show that broadcast with $h$ compliant
and $f$ faulty parties can be implemented using \cast{k} channels if and
only if $2f<kh$.
 
 %\footnote {An earlier preprint of our result, with a 3 point rounding
 % gap between bounds, (and possibly its future improved versions)
 % is available at \cite{CLM}.}

 \section {Definitions and Results}

\begin{dfn}\label{cast} A  {\em \cast{k} channel} is a primitive for
authenticated reliable communication to $k$ parties.
 To use it, one party, the {\em sender}, selects $k$ {\em recipients},
and a message $m$. Each recipient gets $m$, as well as the identities of
the sender and the other recipients. \end{dfn}

\begin{dfn}\label{prot} A {\em protocol} is an algorithm used in {\em
rounds} by several communicating parties. Each party starts with an
input appended with its and other parties' identities. At each round,
parties can \cast{k} messages to be used by the recipients as inputs for
the next round. Besides the algorithm, the interaction is affected by
the {\em Adversary} who selects the initial inputs of all parties and
assigns, possibly with restrictions, their {\em loyalties}, \ie chooses
a subset of {\em faulty} parties and replaces their {\em communications}
(inputs, messages, and outputs) by data of its choice.\end{dfn}

 \begin{dfn}\label{byz} {\em Byzantine agreement} is a broadcast
simulating protocol. The party's {\em value} is its output for a
recipient or input for the sender. The protocol succeeds if the values
of non-faulty (compliant) parties are all identical.\end{dfn}

\begin{thr} {\em Byzantine agreement} protocols for $h$ compliant and $f$
faulty parties using \cast{k} channels exist if and only if $2f<kh$.
\end{thr}

\subsection{Broadcast and Consensus} \label{consensus-s}

In a traditional consensus model, each party starts with an input value.
After running a consensus protocol, all compliant parties output values
consistent with each other and with an input of at least one compliant
party. With a compliant majority, consensus is easily shown to be
equivalent to broadcast. To achieve consensus, each party broadcasts its
value to the others who then output the majority value. To broadcast,
the sender sends its input to all parties, who then run a consensus
protocol on the values received.

This equivalence fails when the majority is faulty. The Adversary gives
inputs $0$ and $1$ to an equal number of parties. They all run the
protocol faithfully. The Adversary defeats the consensus by keeping
compliant some parties with different outputs or declaring faulty all
parties whose inputs match the uniform output.

One can generalize the consensus model, by assuming each party to have
not just one input or output value but rather a {\em distribution}, \ie
a value for each $k$-node set he belongs to. All compliant members of
the set get the same input value for it. All output values of compliant
parties must agree with each other and with at least one input of a
compliant party. This model can simulate broadcast after the sender {\em
distributes} his input, \ie \cast{k}s it to all $k$-node sets.

\section {Proof of the Lower Bound}

\subsection {Big Rings and Chains} \label{ring-s}

The Adversary's ability to defy the agreement will depend on having
enough parties to build a big ring. A {\em $(k,h)$-ring} is a set of
$k\!+\!2$ or more clusters of parties where the clusters are arranged in
a cycle, with at least $h$ parties in any two adjacent clusters. These
bounds assure that no message can be sent to all clusters, and all
compliant parties can fit in any two adjacent clusters. Adding up
parties in all pairs of consecutive clusters, we get $(k\!+\!2)h$ or
more, counting each party twice. So, to build the ring, the Adversary
needs $f\!+\!h\ge (k\!+\!2)h/2$ parties, which means $2f\ge kh$. A
$(k,h)$-ring could be opened into a {\em $(k,h)$-chain} by duplicating
all nodes in the sender's cluster S, creating two clusters, S$_0$ and
S$_1$.

\subsection {The Adversary Strategy}

The Adversary defeats any protocol P if $2f\ge kh$. It arranges the
parties into a $(k,h)$-chain, duplicating the sender's cluster S as in
section~\ref {ring-s}. One copy of S will end up being simulated by the
adversary. Both copies play the same part in protocol P, receiving
duplicate messages from other clusters. The two copies of the sender get
opposite inputs. The messages from S are generated by both copies, but
one copy is intercepted by the Adversary as follows.

Each transmission from S misses all parties in at least one other
cluster. Discounting the leftmost such cluster splits the chain into two
sub-chains: C$_0$ and C$_1$. The Adversary will keep compliant two
adjacent clusters, so either C$_0$ or C$_1$ will have no compliant
parties. Thus the Adversary can and does intercept the messages from
S$_m$ to C$_{1-m}$.

With these restrictions on the Adversary, no messages or outputs depend
on its choice of compliant parties. Since the values of the S$_0$ and
S$_1$ copies of the sender differ, there must be parties in adjacent
clusters that disagree on their value. The Adversary defeats the
protocol by choosing the conflicting clusters as compliant, corrupting
all others. \qed

\section {Proof of the Upper Bound}

We describe a Byzantine protocol P$_h$ for $h$ compliant and $f=\ceil
{kh/2}\!-\!1$ faulty parties. It can also run with fewer, $n<h\!+\!f$,
parties, and is still guaranteed to succeed, provided all $d=h\!+\!f\!-
\!n$ missing parties are counted as faulty if the sender $s$ (who will
represent the missing parties) is faulty. P starts with $s$ distributing
its input, and uses the following concept of trust graphs.

\subsection {Trust Graphs} \label{graph-s}

A {\em trust graph} is formed by each party and links pairs of parties
that report consistently inputs received from $s$ by both. ``Sender
clusters'' S$_m$ are added to the graph, each S$_m$ being a clique of
$1\!+\!d$ nodes connected to all recipients who report uniform inputs
$m$.

A {\em pruning} is then conducted as follows. Because the $h$ compliant
parties must form a clique, edges not in cliques can be removed.
 Since cliques are hard to detect, we remove instead (until none left)
edges $(a,b)$ that do not belong to any {\em bi-star} \ie an $h$-node
star with two centers $a,b$ (adjacent to all its nodes).

We use trust graphs to choose agreement values. 
 All compliant parties must be adjacent in the graph.
 If, in their respective graphs, compliant recipients have paths to a
unique sender cluster, they may immediately output its value.

Consider a path connecting nodes in S$_m$ with different $m$, say, $S_0$
and $S_1$. It must have more than $k$ recipients. Otherwise there would
be one \cast{k} received by them all; since parties connected to S$_m$
claim this \cast{k} was $m$, there must be some disagreement along this
path.

One can break S$_0$, S$ _1$ and the recipients into clusters according
to the distance from S$_0$, dropping nodes more distant than S$_1$. They
form a $(k,h)$-chain, since every two consecutive clusters include an
$h$-node bi-star. So, by section~\ref{ring-s}, a trust graph with a path
between S$_0$ and S$_1$ implies $2f\ge kh$.

\subsection {The Protocol}

Each party $i$ distributes all messages $M_i$ it received from $s$.
 Then all parties except $s$ run P$_h$ recursively to agree on $M_i$ and
form trust graphs based on the agreed $M_i$'s. Let $n$ be the minimal
number of parties for which the guarantee for P$_h$ can fail. Then the
agreement on $M_i$ succeeds unless $s$ is compliant and $i$ faulty.

Thus, the compliant parties always form a clique, and if $s$ is not
among them, the graphs are identical. Then each party with a path to
S$_m$ outputs $m$, or $0$, if no such paths exist. The agreement can
fail only if a path connects both S$_m$. As per Section~\ref{graph-s},
this contradicts $2f<kh$. \qed

 %\section {Acknowledgments}


\newpage\begin{thebibliography}{99}

 \bibitem[STOC]{stoc} {\em
 Proceedings of the Annual ACM Symposium on Theory of Computing.}

{{\bibitem{A} D. Angluin. Local and Global Properties in Networks of
Processors. [STOC], 1980. pp.~82-93.}}

 \bibitem{BG} D. Beaver, S. Goldwasser. Multi Party Fault Tolerant
Computation with Faulty Majority Based on Oblivious Transfer.
 {\em Proceedings of the Annual IEEE Symposium on Foundations of
Computer Science,} 1989, pp.~468-473.

 \bibitem{BGW} M. Ben-Or, S. Goldwasser, A. Wigderson. Completeness
Theorems for Non-cryptographic Fault-tolerant Distributed Computation.
[STOC], 1988, pp.~1-10.

\bibitem{CLM} Jeffrey Considine, Leonid A. Levin, David Metcalf.
Byzantine Agreement with Bounded Broadcast. Preprint at
http://arXiv.org/abs/cs.DC/0012024 , 2000.

{\bibitem{FLM} M. Fischer, N. Lynch, M. Merritt. Easy Impossibility
Proofs for Distributed Consensus Problems. {\em ACM Symposium on
Distributed Computing}, 1985, pp.~59-70.}

 \bibitem{FM} Matthias Fitzi, Ueli Maurer. From Partial Consistency to
Global Broadcast. [STOC], 2000, pp. 494-503.

{\bibitem{FW} M. Franklin, R. Wright. Secure Communication in Minimal
Connectivity Models. {\em Advances in Cryptology (Eurocrypt)}, 1998,
pp.~346-360.}

{\bibitem{FY} M. Franklin, M. Yung. Secure Hypographs: Privacy from
Partial Broadcast. [STOC], 1995, pp.~36-44.}

{\bibitem{GGL} O. Goldreich, S. Goldwasser, N. Linial. Fault-tolerant
Computation in the Full Information Model. {\em SIAM J. Computing},
27(2), 1998, pp.~506-544.}

\bibitem{GMW} O. Goldreich, S. Micali, A. Wigderson. How to Play Any
Mental Game, or A Completeness Theorem for Protocols with Honest
Majority. [STOC], 1987, pp.~218-229.

\bibitem{GL} Shafi Goldwasser, Leonid A. Levin. Fair Computation of
General Functions in Presence of Immoral Majority. {\em Annual CRYPTO
Conference (IEEE/LACR)}, Santa Barbara, August 1990 (Proceedings 1991).

 \bibitem{DFFLS} Danny Dolev, Michael J. Fischer, Rob Fowler, Nancy A.
Lynch, and H. Raymond Strong.
 An efficient algorithm for Byzantine agreement without authentication.
 {\em Information and Control}, 52(3):257-274, 1982.

 \bibitem{PSL} M. Pease, R. Shostak, L. Lamport. Reaching agreement in
the presence of faults. {\em J. Comput. Mach.} 27:121-169, 1980.

 \bibitem{RB} T. Rabin, M. Ben-Or. Verifiable Secret Sharing and
Multi-party Protocols with Honest Majority. [STOC], 1989, pp. 73-85.

{{\bibitem{WD} Y. Wang, Y. Desmedt. Secure Communication in Multicast
Channels: The Answer to Franklin and Wright's Question. {\em Advances in
Cryptology (Eurocrypt)}, 1999, pp.~446-458.}}

\end{thebibliography}
  \end{document}